\documentclass[journal=jacsat,manuscript=article, email=false]{achemso}

\usepackage[version=3]{mhchem} 
\usepackage[table,xcdraw]{xcolor}

\newcommand*{\editor}[2]{%
  \expandafter\newcommand\csname #1note\endcsname[1]{%
    \textcolor{#2}{(\textbf{#1:} \it ##1)}}%
  \expandafter\newcommand\csname #1\endcsname[1]{%
    \textcolor{#2}{##1}}%
  \expandafter\newcommand\csname #1cancel\endcsname[1]{%
    \textcolor{#2}{\sout{##1}}}%
  \expandafter\newcommand\csname #1change\endcsname[2]{%
    \textcolor{#2}{\sout{##1} ##2}}%
  \newenvironment{#1text}{\color{#2}}{\color{black}}
}

\usepackage{xcolor}
\usepackage{comment}
\definecolor{Teal}{rgb}{0.00,0.50,0.50}
\definecolor{rnd}{rgb}{0.50,0.00,0.50}
\definecolor{Gray}{gray}{0.9}

\editor{FP}{Teal}
\editor{CS}{rnd}

\title{State- and momentum-dependent nonlinear Stark effect of interlayer excitons in bilayer WSe$_2$}

\author{Cem Sevik}
\affiliation{Department of Physics and NANOlight Center of Excellence, University of Antwerp, Groenenborgerlaan 171, B-2020 Antwerp, Belgium}
\email{cem.sevik@uantwerpen.be}
\author{Engin Torun}
\affiliation{Department of Physics and NANOlight Center of Excellence, University of Antwerp, Groenenborgerlaan 171, B-2020 Antwerp, Belgium}
\author{Milorad V. Milo\v{s}evi\'c}
\affiliation{Department of Physics and NANOlight Center of Excellence, University of Antwerp, Groenenborgerlaan 171, B-2020 Antwerp, Belgium}
\author{Fulvio Paleari}
\affiliation{Centro S3, CNR-Istituto Nanoscienze, I-41125 Modena, Italy}
\email{fulvio.paleari@nano.cnr.it}

\begin{document}



\begin{abstract}
Interlayer excitons in van der Waals heterostructures offer rich collective phases, prospective optoelectronic applications, and versatile tunability, where control by electronic means is particularly relevant and practical. Here, in the case of bilayer WSe$_2$, we reveal how layer localization of excitons governs their response to an external electric field. Using Many-Body Perturbation Theory, we calculate the exciton dispersion for different stacking symmetries under applied electric field and/or strain, in order to map the landscape of competing low-energy excitons in four distinct finite-momentum valleys. While intralayer excitons are not affected by the electric field, some interlayer ones exhibit a nonlinear Stark shift that becomes linear after a critical threshold. The degree of nonlinearity is a direct measure of the layer hybridization of the electronic subcomponents of the exciton. Our findings explain the peculiar Stark-shift regimes observed in recent experiments, the nature of (anti)symmetric spectral shifts around zero field, and the sensitivity of dipolar excitons to external perturbations, all highly relevant to their further applications in excitonic condensates, optoelectronics devices and quantum emitters.

\end{abstract}

\maketitle
The layered two-dimensional transition metal dichalcogenides (TMDs) can host spatially separated excitons (interlayer, IL) where the Coulomb-bound electron and hole tend to lie on separate layers\cite{jiang2021interlayer}. 
Recent studies have revealed the exceptionally long lifetimes, up to a microsecond, of IL excitons in bilayer/heterobilayer TMDs\cite{jiang2018microsecond, TMDL6, TMDL7, TMDL8,TMDL9, TMDL10, TMDL11,TMDL12, TMDL1, TMDL2, TMDL3, TMDL4, TMDL5, huang2022spatially,WangNL}, due to the weak overlap between the electron and hole wavefunctions. These discoveries have sparked a growing interest in these systems as platforms to investigate emerging frontier quantum many-body phenomena such as quantum-confined Stark effect\cite{WangNL, leisgang2020giant}, exciton Bose-Einstein condensation\cite{eisenstein2004bose, wang2019evidence, fogler2014high}, super-fluidity at high temperature\cite{wang2019evidence}, and excitonic insulator states\cite{ma2021strongly,rivera2018interlayer}. Among these materials, bilayer (BL) WSe$_2$ is particularly noteworthy for investigating IL excitons with finite center-of-mass (COM) momentum, leading to momentum-indirect, phonon-assisted spectral emission peaks\cite{WangNL,huang2022spatially}.

Lately, Huang \textit{et al.}\cite{huang2022spatially} conducted a rigorous experimental investigation into these excitons using electric field-dependent photoluminescence (PL) measurements (the field $E_z$ being along the out-of-plane direction of the WSe$_2$ BL). 
Their findings revealed the formation of several IL exciton states attributed to electron-hole transitions taking place across distinct valleys, such as $\Lambda$-K and $\Lambda$-$\Gamma$. 
This improved upon the original measurement of the quantum-confined Stark effect by Wang \textit{et al.}\cite{WangNL}, where only one IL exciton type was assumed. 
Notably, the PL measurements confirm the slight energy differences among band gaps between different valleys, while also demonstrating two particularly significant effects: (i) an apparent crossing in the lowest-energy exciton levels induced by the electric field and (ii) the signature of two distinct regimes in the quantum-confined Stark effect undergone by some IL excitons - a nonlinear Stark shift at small $E_z$ transitioning into a linear Stark shift at large $E_z$. 
Furthermore, they successfully measured the electric-field\textit{-independent} intralayer (in-plane, IP) direct exciton energy, this state originating from vertical electron-hole transitions around the K point. 
As already mentioned, the quantum-confined Stark effect involving IL excitons had previously been demonstrated for the same material by different research groups within a similar exciton energy window\cite{Scuri1,chen2019band, WangNL}. 
In fact, Wang \textit{et al.}\cite{WangNL} made similar predictions regarding the presence of two distinct regimes for the Stark shift and attributed it to the doped carriers requiring a threshold electric field in order to fully localize within one layer. 
Lately, Tagarelli \textit{et al.}\cite{tagarelli23} conducted analogous measurements, focusing on characterizing the properties of \textit{hybrid}-type excitons (i.e., having both IL and IP character). 
They reported a significant influence of the applied out-of-plane electric field on the degree of hybridization of the excitons, which was maximum at zero field and minimum (meaning almost pure IL states) at high field, which allows for controlling the degree of deviation from the diffusive regime in exciton transport measurements.

On the theoretical side, while no predictive assessment of these effects currently exists for BL WSe$_2$, two recent studies are available on \textit{zero-momentum} excitons on different TMD systems\cite{deilmann2024quadrupolar,jasinski2024quadrupolar}, proposing a specific framing of the electric-field dependent states as ``quadrupolar'' (nonlinear shift) excitons as opposed to their ``dipolar'' counterparts (linear shift).
While earlier first-principles-based methodologies hold promise for addressing these issues\cite{bernardi2017optical,vitale2021flat}, the subtle variations in exciton relative energies demand calculations of very high precision, and including finite COM excitons originating from nonvertical electron-hole transitions. 

To address this issue, in this work we: (i) provide a clear demonstration of the nonlinear and linear regimes of the Stark shift for momentum-dark, hybrid IL-IP excitonic states in BL WSe$_2$; (ii) demonstrate that these effects are single-particle in nature and depend on the degree of layer hybridization of the electrons forming the exciton complexes, which can be controlled by the external electric field;  (iii) include a strain-dependent study of the excitonic properties, combined with external gating, as a way to retain predictive capabilities in first-principles simulations when investigating properties that are essentially sample-dependent and thus may vary not only as a consequence of theoretical approximations, but also because of experimental sample preparation (such as local strain and substrate screening effects); (iv)
we link the asymmetric E$_z$ response to the stacking type, proposing that the sample measured by Tagarelli \textit{et al.} \cite{tagarelli23} was in a mirror-symmetric stacking (AB) rather than the assumed centrosymmetric stacking (AA$^\prime$). 

\section{Results and discussion}
\begin{figure*} [htbp]
    \centering
        \includegraphics[width=16cm]{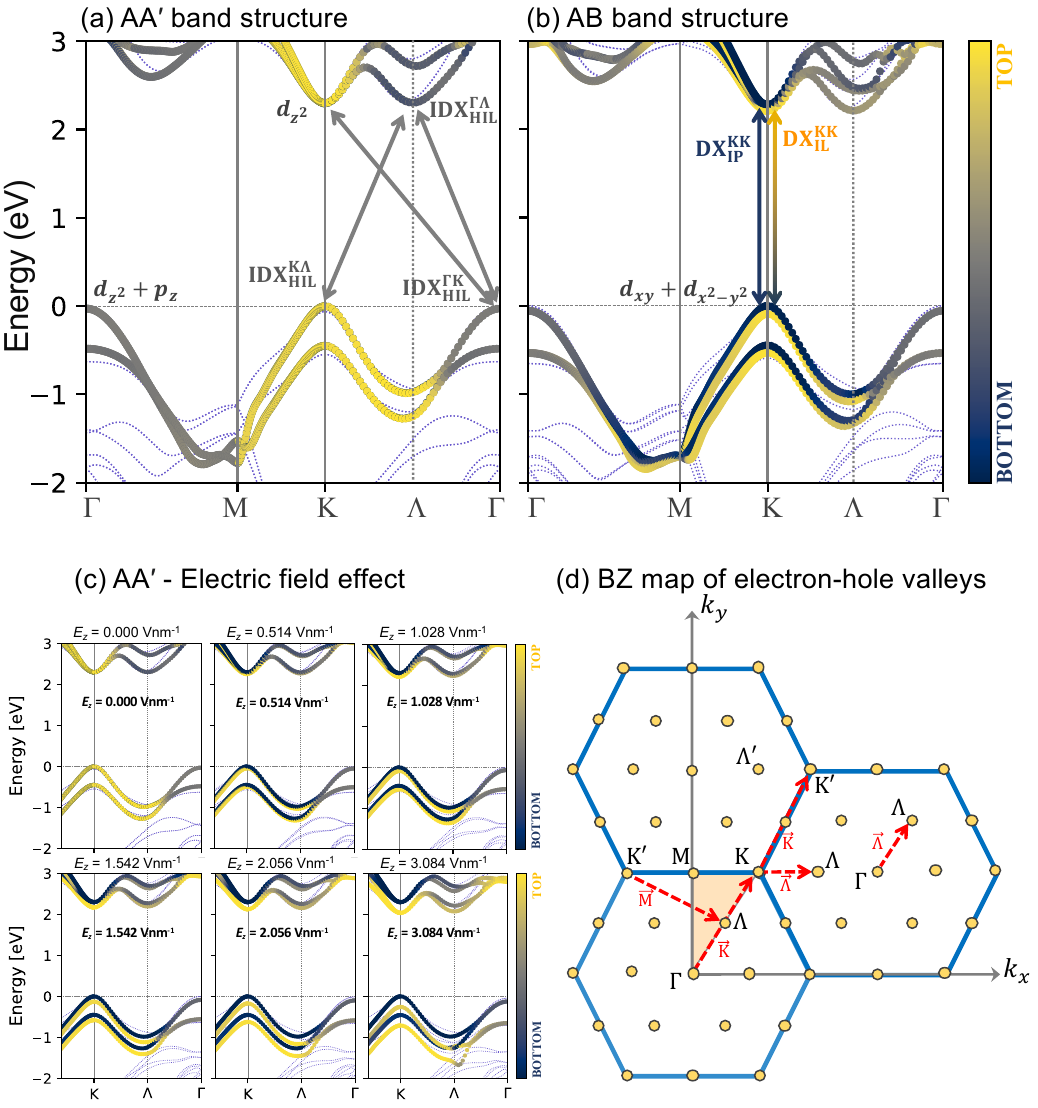}
            \caption{The calculated DFT (dashed lines) and G$_0$W$_0$ (colored circles) band structures for (a) AA$^{\prime}$ and (b) AB stacked bilayer WSe$_2$. The color bar represents the contribution of individual layers to the corresponding states. The DFT bandgap at the K point is shifted to the G$_0$W$_0$ bandgap value for comparison. The high symmetry points in the hexagonal BZ and the most prominent single particle transition for each exciton (X) are shown in the figures as well. (c) Field-dependent variations in the band structure of the AA$^\prime$ system,  including changes in layer localization. (d) Sketch of the momenta (red arrows and labels) associated to each low-energy electronic transition between band extrema, corresponding to the valleys of the exciton dispersion curves such as in Fig. \ref{fig3}.}
		\label{fig2}
\end{figure*}
Our study was conducted in the framework of first-principles, post-Density Functional Theory (DFT), Many-Body Perturbation Theory methodologies. 
In particular, we adopted the G$_0$W$_0$ approximation in order to correct the DFT band energies and the Bethe-Salpeter Equation (BSE) to calculate the finite-momentum bound exciton dispersion curves\cite{Strinati1988,Martin2016}.
The technical details about the theoretical-computational framework, including numerical parameters and tests conducted, along with structural information and explanation of the excitonic states labeling, are presented in the Supplementary Materials File (SMF). 
The SMF includes the following additional Refs.~\citenum{perdew1996generalized,hamann2013optimized,van2018pseudodojo,RodriguesPela2024,grimme2006semiempirical,barone2009role,muhammad2021structural,nguyen2015dispersion,wilson2017determination,sohier2017density,marini2009yambo,sangalli2019many,Rozzi2006,guandalini2023efficient,hedin1970effects,onida2002electronic,Godby1989,Oschlies1995,smith2017interacting,jia2019charge,liu2020electrically,zhao2013origin,mahmoudi2023electronic,andersen2021excitons,zhao2013evolution,arora2020stacking,finteis1997occupied,nakamura2020spin,agnoli2018unraveling}.

\textit{WSe$_2$ homobilayer at zero field and zero strain.}
The layer-projected band structures of AA$^{\prime}$ (relatively 180$^\circ$ rotated layers) and AB-stacked (0$^\circ$ stacking) bilayer WSe$_2$ are shown in Fig.~\ref{fig2}. 
At the DFT level, the valence band maximum (VBM) at the K point is 61 meV (AA$^{\prime}$) and 40 meV (AB) \textit{lower} in energy than the $\Gamma$-point maximum, resulting in an indirect band gap. 
However, after G$_0$W$_0$ corrections, the K-point VBM becomes 26 meV (AA$^{\prime}$) and 56 meV (AB) \textit{higher} than the $\Gamma$-point maximum. 
Angle-resolved photoemission spectroscopy (ARPES) measurements\cite{PhysRevB.91.041407, zhao2013evolution, agnoli2018unraveling, yuanNL, wilson2017determination, zhang2016electronic} on samples synthesized or fabricated using various methods also identify the K-valley as the highest energy level, with reported $\Gamma$-K energy differences of 140, 80, 50, and 40 meV. 
For both stacking arrangements, the indirect K-$\Gamma$ band gap at the DFT level transitions to a direct K-K band gap at the G$_0$W$_0$ level. 
The corresponding K-K band gap values are 1.135 eV and 2.302 eV for AA$^{\prime}$, and 1.073 eV and 2.207 eV for AB stacking, at the DFT and G$_0$W$_0$ levels, respectively. 
Notably, the energy difference between the direct (K-K) and indirect ($\Lambda$-K) transitions at the G$_0$W$_0$ level is minimal -- only 3 meV and 1 meV for the AA$^{\prime}$ and AB stackings, respectively -- indicating these systems are quasi-direct at the single-particle level (in terms of photoemission energies).

The calculated G$_0$W$_0$ single-particle band features correspond to four competing excitonic valley minima at center-of-mass momenta: $\mathbf{Q}=0$, $\mathbf{Q}=\boldsymbol{\Lambda}$, $\mathbf{Q}=\mathbf{K}$, and $\mathbf{Q}=\mathbf{M}$. 
Excitons with the same COM momentum (i.e., in the same valley) may arise from single-particle transitions involving different regions of the Brillouin zone: $\Gamma$-K and K-K$^\prime$ both yield $\mathbf{Q}=\mathbf{K}$, while K-$\Lambda$ and $\Gamma$-$\Lambda$ result in $\mathbf{Q}=\boldsymbol{\Lambda}$. 
The $\mathbf{Q}=\mathbf{M}$ valley originates from degenerate K$^\prime$-$\Lambda$/K-$\Lambda^\prime$ transitions. 
\begin{figure*} [htbp]
    \centering
        \includegraphics[width=16cm]{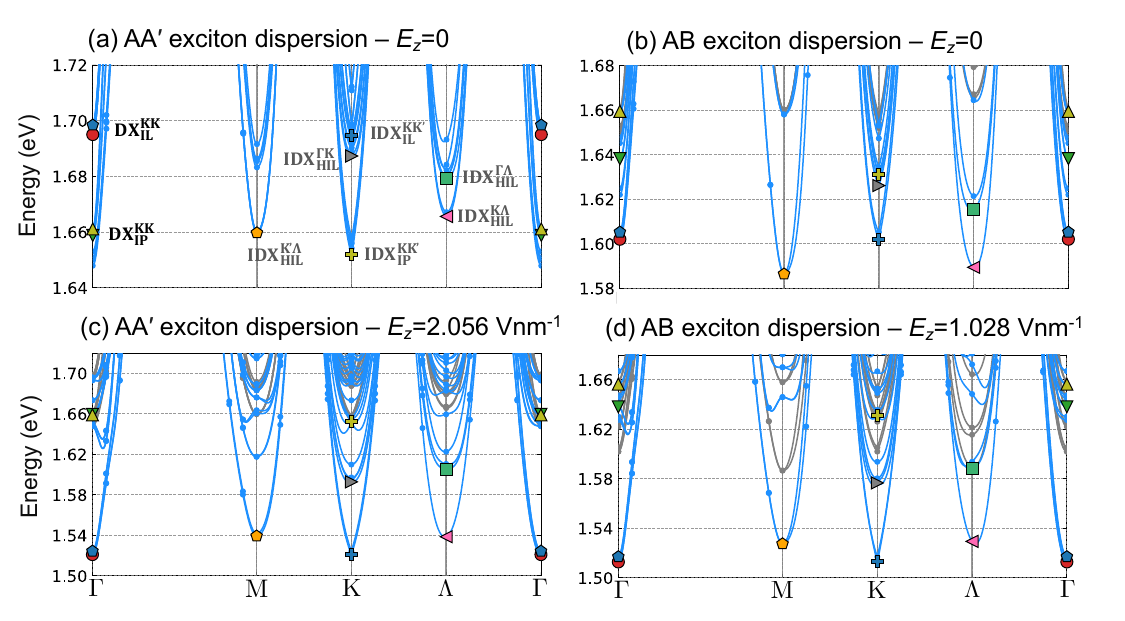}
            \caption{Exciton dispersion curves (blue) of bilayer WSe$_2$ for (a) AA$^{\prime}$ stacking at $E_z=0$, (b) AB stacking at $E_z=0$, (c) AA$^{\prime}$ with $E_z=2.056$ V/nm (with zero-field branches in gray), (d) AB with $E_z=1.028$ V/nm (again zero-field branches in gray). The excitonic state whose Stark shift was studied are labeled consistently with Fig.~\ref{fig4}.}
		\label{fig3}
\end{figure*}
The detailed definition of these direct and indirect excitons (DX and IDX), along with layer localization nature -- IP, IL, hybrid-interlayer (HIL) -- are presented in Fig. \ref{fig2}. 
The corresponding exciton dispersion curves, obtained as described in the SMF at zero field, are plotted in Fig.~\ref{fig3}(a) for AA$^\prime$ stacking and Fig.~\ref{fig3}(b) for AB stacking. 
For AA$^\prime$ stacking, which is the focus of most available experimental studies, the lowest-lying exciton valleys correspond to the $\Gamma$ and K points. 
The dispersion clearly depicts the order of the zero-momentum in-plane excitons, DX$_{\textrm{IL}}^{\mathrm{KK}}$ and DX$_{\textrm{IP}}^{\mathrm{KK}}$, as well as the finite-momentum hybrid interlayer excitons IDX$_{\textrm{HIL}}^{\mathrm{K}\Lambda}$, IDX$_{\textrm{HIL}}^{\Gamma\Lambda}$, and IDX$_{\textrm{HIL}}^{\Gamma\mathrm{K}}$. 
Reported PL measurements on the energy of the lowest direct exciton peak, varying between 1.62 and 1.76 eV across different sample configurations, such as as-grown, transferred, and hBN-encapsulated WSe$_2$,\cite{zhao2013evolution, WangNL, chen2019band, jia2019charge, liu2020electrically, zhao2013origin, mahmoudi2023electronic, andersen2021excitons, huang2022spatially, Scuri1, arora2020stacking, debnath2022tuning, barman2022twist} align with our results. 

The energy ordering between IP and IL states is reversed in the AB case. 
The very high degree of segregation of electrons and holes at the K point in different layers in the DX$_{\textrm{IL}}^{\mathrm{KK}}$ case results in weaker Coulomb binding.
In addition, the change in interlayer interaction due to the shifted atomic planes causes a significant splitting of the IP states.
We point out that while the AB system loses inversion symmetry, it retains the $C_3$ rotations of the $C_{3v}$ group, so that each IP and IL exciton state remains doubly degenerate ($E$ representation) like in the AA$^\prime$ case.
Although the electronic transitions K-K and $\Lambda$-K have nearly the same energy, the IDX$^{ \textrm{K}\Lambda}_{\textrm{HIL}}$ exciton has a 12 meV stronger binding energy than the direct IL exciton, making this system a rare example of (quasi-)direct single-particle band gap (probed by photoemission) featuring an ``indirect'' neutral excitation spectrum (probed by linear optical absorption and emission).
The experiment by Tagarelli \textit{et al.} \cite{tagarelli23} supports our findings, as the energetically lowest state was reported to be the same HIL exciton, possessing higher radiative decay rates and effective dipole lengths.
The reason why we consider their sample to be in the AB -- rather than AA$^\prime$ -- stacking will be clarified below.
Note that, due to the asymmetric hybridization level of the constituting electron-hole pair (layer-segregated hole at K and layer-delocalized electron at $\Lambda$), the finite-$\mathbf{Q}$ IDX$^{ \textrm{K}\Lambda}_{\textrm{HIL}}$ state is referred to as a HIL exciton. 
We will see that hybrid-IL excitons in bilayer WSe$_2$ correspond to ``quadrupolar'' excitons displaying nonlinear Stark shifts.
For both stacking arrangements, the exciton binding energies were calculated between 600 - 650 meV (we define the binding energy as the difference between the minimum single-particle transition and the exciton energy \textit{with the same momentum transfer $\mathbf{Q}$}).
We note that the literature showcases a variety of experimental estimations, spanning both lower and higher values \cite{hanbicki2015measurement, BEBE-1, BEBE-2, BEBE-3}. 

\begin{figure} [!ht]
		\centering
		\includegraphics[width=15cm]{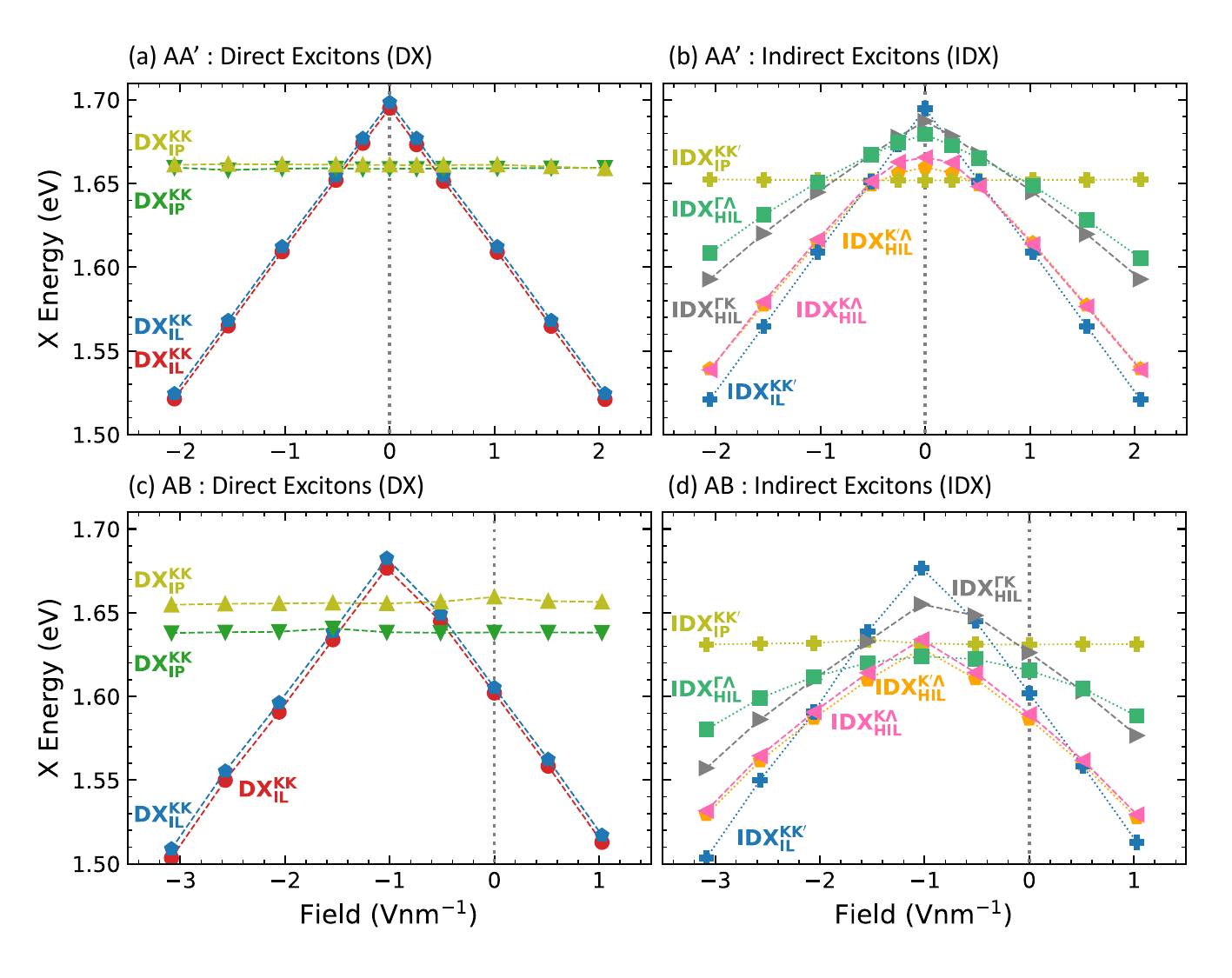}
		\caption{Quantum-confined Stark effect. Variation of the calculated exciton energies with respect to the external out-of-plane electric field $E_z$  for (a) AA$^{\prime}$ and (b) AB stacked BL WSe$_2$. X$_{IP}$: the calculated direct IP exciton energy corresponding to the lowest-lying dominant absorption peak, X$_{\Lambda\Gamma}$: the calculated indirect minimum energy exciton between $\Lambda$-$\Gamma$, X$_{\textrm{K}\Gamma}$: the calculated in-direct minimum energy exciton between K-$\Gamma$, X$_{\Lambda\textrm{K}}$: the calculated in-direct minimum energy exciton between $\Lambda$-K, and X$_{IL}$: the calculated direct IL minimum energy exciton.}
		\label{fig4}
  \end{figure}


\textit{Field dependence of the exciton valleys.}
In Fig.~\ref{fig3}(c) and (d) we illustrate the change in the exciton dispersions when an electric field $E_z$ along the out-of-plane direction is applied.
We immediately notice that the IL states now become the lowest ones in the AA$^\prime$ case ($E_z=2$ V/nm), while in the AB case ($E_z=1$ V/nm) now the minima of the exciton dispersion also lie at the $\Gamma$ and K valleys.
Thus, the exciton bandwidth undergoes both internal state reordering and an overall redshift.
In order to investigate this further, we studied the exciton energies variation as a function $E_z$ -- see Fig.~\ref{fig4}, the labeling of the excitonic states being consistent with Fig.~\ref{fig3} -- for both stackings\footnote{We have selected the lowest-energies excitonic states in the various $\mathbf{Q}$-valleys displaying dipolar character (i.e., changing their energy with $E_z$) and included the in-plane excitons at $\mathbf{Q}=0$ for comparison. The exciton energy level crossings with the electric fields were unambigouously identified by plotting the excitonic wave functions in real space at different field values for the levels involved.}.
The left plots show the behaviour of the bright $\mathbf{Q}=0$ excitons for both stackings, while the right plots are concerned with the the finite-Q states in the other excitonic valleys. 
Due to the breaking of inversion symmetry along the $z$-direction, which splits the exact band alignment of the constituent layers as shown in Fig.~\ref{fig2}(b), the electric field response of the AB stacking is not symmetric around zero field but centered around -1 Vnm$^{-1}$.\footnote{Please note that the field is applied along the 60 {\AA} long periodic supercell. Therefore, -1 Vnm$^{-1}$ is a relative value. If we change the cell size, the required field to merge the bands corresponding to different layers into a degenerate state, such as in the zero-field AA$^\prime$ configuration, will change.}
In fact, Tagarelli \textit{et al.} attributed the asymmetric field dependence observed in their experiment to intrinsic doping in the WSe$_{2}$ BL\cite{tagarelli23}. 
However, this observed feature follows a trend similar to our calculation results, as seen in Fig.~\ref{fig4} (c) and (d). 
Therefore, we propose that the sample in this experiment was actually an AB-stacked WSe$_2$ BL.
We therefore remark that the asymmetry in the Stark shift may provide a field-mediated determination of competing stacking arrangements in TMD layered structures. 

Remarkably, our predictions regarding the impact of $E_z$ on exciton energies strongly agree with the recent experimental findings about the quantum-confined giant Stark effect\cite{huang2022spatially,WangNL,wang2019evidence, altaiary2022electrically}. 
As depicted in Fig. \ref{fig4} (a) and (b), the phenomenology of the Stark effect varies significantly depending on the specific exciton type. 
The distinct behaviours primarily depend on the degree of layer hybridization of the electron-hole pairs which make up the various exciton states. 
The behaviour of the K and $\Lambda$ conduction band valleys with respect to the external field is depicted for the AA$^{\prime}$ in Fig.~\ref{fig3}(c), showing, among other things, a linear decrease of the direct (K-K) band gap with the field.
Because of field-induced symmetry breaking, the electronic states acquire a field-dependent splitting, which can combine with the zero-field spin-orbit coupling (SOC) induced splitting, as observed also in mirror-symmetric trilayers\cite{deilmann2024quadrupolar}.
In particular, the states which are layer-degenerate at zero field all split into layer-segregated electron and holes at finite field (notably, there is an additional SOC splitting in the valence K-valley at zero field).
The field-dependent splitting in the conduction $\Lambda$ valley is weaker than at K (note that here the states are already partially SOC-split, a negligible effect in the conduction K valley).
For zero field, the $\Lambda$ electrons are delocalized on both layers, while for large fields, they split in four levels (combination of SOC+field splitting) two of them becoming more and more localised on the top and bottom layers.

Let us now discuss the behaviour of the exciton states more in detail.
Going back to Fig.~\ref{fig4}, we notice immediately that the $\mathbf{Q}=0$ intralayer (in-plane) exciton DX$^{\mathrm{KK}}_{\textrm{IP}}$ is unaffected by $E_z$, in agreement with experimental observations \cite{tagarelli23,klein2016stark,wang2017probing,WangNL,chen2019band,huang2022spatially}, since it lacks an out-of-plane dipole moment. 
In the case of the $\mathbf{Q}=0$ interlayer direct excitons (DX$^{\mathrm{KK}}_{\textrm{IL}}$), the Stark shift follows a linear trend, consistent with the expected behavior of the atomic Stark effect and therefore compatible with the modellization as a classical dipole. 
On the contrary, due to the layer hybridization of electrons in the $\Lambda$ valley and holes in the $\Gamma$ valley, the $\mathbf{Q}=\boldsymbol{\Lambda}$ IDX$_{\textrm{HIL}}^{\textrm{K}\Lambda}$, IDX$_{\textrm{HIL}}^{\Gamma\Lambda}$, and IDX$_{\textrm{HIL}}^{\Gamma\textrm{K}}$ excitons all feature partial spatial confinement character in the exciton channel: consequently, they undergo a transition from a non-linear Stark shift at low field to a linear Stark shift at high field, as demonstrated in Fig. \ref{fig4} (a) and (b).
This observation is in full agreement with recent experimental measurements conducted on high-quality $h$-BN encapsulated BL WSe$_2$ devices \cite{tagarelli23,huang2022spatially,WangNL,wang2019evidence, altaiary2022electrically} and reinforce theoretical findings recently reported for $\mathbf{Q}=0$, direct excitons in WS$_2$/MoS$_2$/WS$_2$ trilayers.
Indeed, as seen before from Fig.~\ref{fig3}, the finite-$\mathbf{Q}$ excitons that start off as hybridized become increasingly layer-localized with $E_z$ \cite{ramasubramaniam2011tunable}: the indirect excitons can be regarded as spatially separated excitons at higher fields. 
Only the $\mathbf{Q}=\textrm{K}$ states originating from electronic transition between K and K$^\prime$ points, which share the same wave function character as the direct K-K transitions, behave linearly like the $\mathbf{Q}=0$ IL states.
For instance, in AA$^{\prime}$ stacking case, the estimated layer hybridization ratio of conduction and valence K points (obtained via wave function projection onto atomic orbitals localized on one or the other layers) is almost 1/0 (top/bottom or bottom/top) and does not vary significantly with the field, whereas for $\Gamma$ valence and $\Lambda$ conduction, the ratios, 0.5/0.5 and 0.67/0.33 at $E_z$ = 0 Vnm$^{-1}$, change linearly to the ratios 0.33/0.67 and 0.80/0.20 at $E_z$ = 0.308 Vnm$^{-1}$, see Fig. \ref{fig3}(c).
Since the nonlinear Stark shift can be approximated as a quadratic one at low fields, one could model these bound states as ``quadrupolar'' excitons originating from an effective interaction between fictitious, linearly dispersing ``dipolar'' excitons, yielding the following expression for the field-dependent exciton energies (see Ref.~\citenum{deilmann2024quadrupolar}): $E_X = \sqrt{(p E_z)^2 + t^2}$. Here, $p$ is the effective dipole length and $|t|$ is an effective coupling term expressing the degree of deviation from the dipolar model.
We have fitted our \textit{ab initio} results according to this expression and the results are listed in Table~\ref{tab1} (first two columns).
The values of $p$ are primarily determined by the underlying single-particle band structure of the systems, while $t$ seems to depend both on the band structure and on the details of the microscopic electron-hole interaction kernel $\hat{\Xi}(\mathbf{Q})$ (see SMF for definition). 
We see that $p$ is around 0.08 nm for for the K-K/K-K$^\prime$ transitions, while nonlinear states tend to have similar dipole lengths if the hole components originate around either the K/K$^\prime$ (W-$d_{xy}+d_{x^2-y^2}$ orbital character) or the $\Gamma$ (W-$d_{z^2}$$+$Se-$p_z$) points, respectively. 
Interestingly, we find an extremely strong deviation from the dipolar case for the $\mathbf{Q}=\boldsymbol{\Lambda}$ case in the AB stacking (green squares in Fig.~\ref{fig4}(d)), with $t\sim 2\times 10^4$ meV, three orders of magnitude above the rest.
All this underscores the fact that the different orbital characters of the electronic wave functions making up the different excitons states largely determines their behavior under electric field, determining both intervalley and intravalley exciton energy relative shifts.\footnote{We also note that the experimentally estimated values of $p$ from Refs.~\citenum{huang2022spatially} and~\citenum{tagarelli23} are systematically larger than in our results, ranging from $0.14$ to $0.45$ nm. These estimations were obtained fitting the signal from phonon-assisted exciton recombination replicas in field- and energy-dependent photoluminescence maps. We speculate that, beside fitting error bars, the presence of a substrate with its enhanced electronic screening effects could result in a significant impact on these parameters, and the same would be true for residual doping, while we are simulating a freestanding, perfectly neutral bilayer.}

\renewcommand{\arraystretch}{1.55}
\begin{table*}[htbp] 
\centering
\begin{tabular}{c |c@{\hspace{0.5em}}c|c@{\hspace{0.5em}}c|c@{\hspace{0.5em}}c|c@{\hspace{0.5em}}c|}
\cline{2-9} 
& \multicolumn{2}{c|}{\textbf{AA}$^\prime$} 
& \multicolumn{2}{c|}{\textbf{AB}} 
& \multicolumn{2}{c|}{\textbf{AA}$^\prime$ planar comp.} 
& \multicolumn{2}{c|}{\textbf{AA}$^\prime$ stacking comp.} \\ 
\rowcolor{Gray}
& $p$ & $t$ & $p$ & $t$ & $p$ & $t$ & $p$ & $t$ \\ \hline \hline
$\mathbf{DX_{IL}^{KK}}(\mathbf{Q}=\boldsymbol{\Gamma})$ & 0.085 & -- & 0.083 & -- & 0.082 & -- & 0.080 & -- \\ 
\rowcolor{Gray}
$\mathbf{IDX_{IL}^{KK^{\prime}}}(\mathbf{Q}=\mathrm{\textbf{K}})$ & 0.084 & -- & 0.084 & -- & 0.082 & -- & 0.079 & -- \\ 
$\mathbf{IDX_{HIL}^{\Gamma K}}(\mathbf{Q}=\mathrm{\textbf{K}})$ & 0.048 & 8 & 0.051 & 20 & 0.046 & 7 & 0.044 & 2 \\ 
\rowcolor{Gray}
$\mathbf{IDX_{HIL}^{K^{\prime}\Lambda}}(\mathbf{Q}=\mathrm{\textbf{M}})$ & 0.082 & 55 & 0.056 & 20 & 0.081 & 58 & 0.072 & 38 \\ 
$\mathbf{IDX_{HIL}^{K\Lambda}}(\mathbf{Q}=\boldsymbol{\Lambda})$ & 0.079 & 39 & 0.057 & 17 & 0.074 & 32 & 0.069 & 20 \\ 
\rowcolor{Gray}
$\mathbf{IDX_{HIL}^{\Gamma\Lambda}}(\mathbf{Q}=\boldsymbol{\Lambda})$ & 0.042 & 20 & 0.600 & 19331 & 0.040 & 18 & 0.036 & 10 \\ \hline \hline
\end{tabular}
\caption{Comparison table showing the fitted values of the dipole effective lengths $p$ in nm, and the effective coupling $t$ in meV. The first two columns refer to pristine AA$^\prime$ and AB stackings. The third and fourth column show the AA$^\prime$ stacking under compression in the layer plane (0.5\% of the in-plane lattice parameter) and along the stacking axis (1\% of the interlayer distance), respectively.}\label{tab1}
\end{table*}


\begin{figure*} [htbp]
    \centering
        \includegraphics[width=15cm]{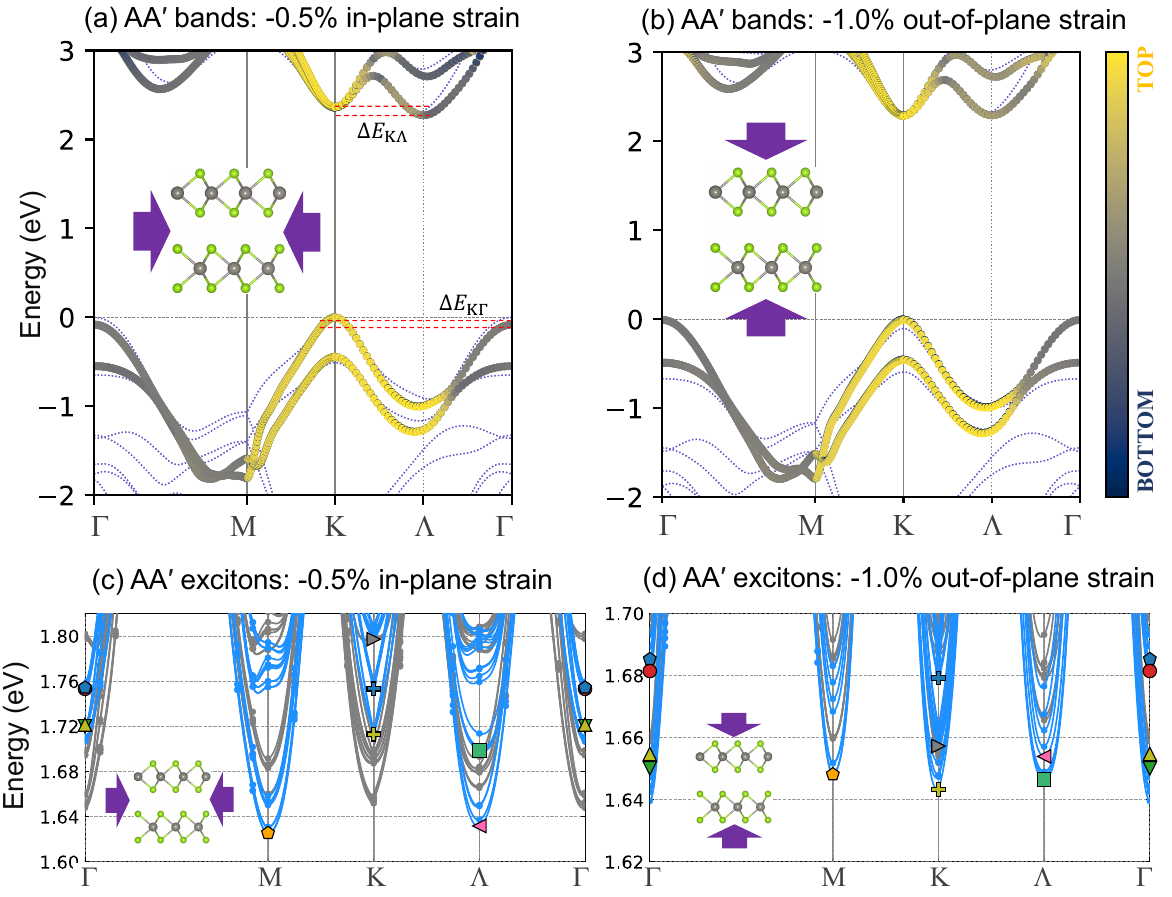}
            \caption{[(a) and (b)] The calculated DFT (dashed lines) and G$_0$W$_0$ (colored circles) band structures for AA$^\prime$-stacked bilayer WSe$_2$ for (a) planar compressive strain (0.5\% of the planar alttice parameter) and (b) compression along the stacking axis (1\% of the interlayer distance). The color bar represents the contribution of individual layers to the corresponding states. [(c) and (d)] Calculated exciton dispersion curves (in blue) corresponding to the (a) and (b) band structures, respectively. The gray curves correspond to the equilibrium structure.}
		\label{fig5}
\end{figure*}
  
\begin{figure}[ht!]
		\centering
	\includegraphics[width=15cm]{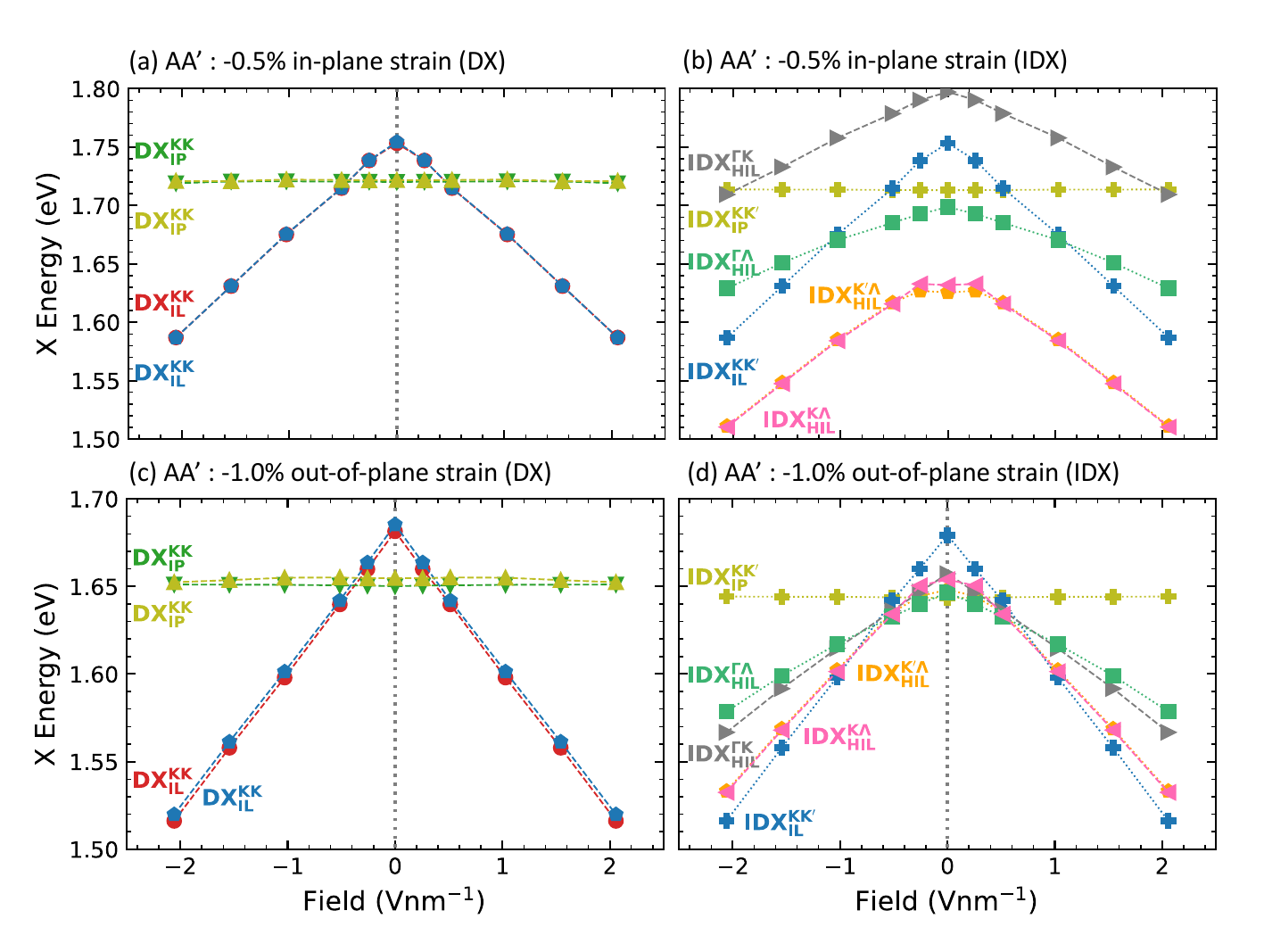}
		\caption{The calculated exciton energies with respect to the external out-of-plane electric field ($E_z$) for (a) in-plane and (b) out-of-plane compressive strain in AA$^{\prime}$ BL WSe$_2$. The exciton state labelling is consistent with Figs.~\ref{fig3},\ref{fig4} and \ref{fig5}.}
		\label{fig6}
  \end{figure}

\textit{Strain dependence of the exciton levels.} 
Although the calculated exciton Stark shifts for different types of excitons closely match experimental observations, the ground state exciton in our calculations at zero field was identified as DX$^{\mathrm{KK}}_{\textrm{IP}}$ due to the previously mentioned direct band-gap nature of the AA$^{\prime}$ stacking. 
To further investigate this issue, we examined bilayer WSe$_2$ subjected to 0.5\% in-plane (IP) (0.0167 {\AA} reduced in-plane lattice constant) and 1.0\% out-of-plane (OP) (0.0641 {\AA} reduced interlayer separation) compressive strain.
The respective effects on the band structures are shown in Fig.~\ref{fig5}(a) and (b). 
In the 0.5\%-IP case, the material shifted to an indirect $\Lambda$-K band gap at 2.271 eV.
The conduction valley energy difference $\Delta\textrm{E}_{\textrm{K}\Lambda}$ is 89 meV, while the valence valley difference $\Delta \textrm{E}_{\textrm{K}\Gamma}$ is 82 meV.
The effect on the bound exciton dispersion is quite clear from Figs.~\ref{fig5}(c), where the gray curves represent the unstrained system: the planar compression causes the M and $\Lambda$ valleys to become the lowest in energy, with the ground state exciton now being the $\mathbf{Q}=\boldsymbol{\Lambda}$ IDX$_{\textrm{HIL}}^{\textrm{K}\Lambda}$, with an energy of 1.632 eV, consistent with the experiment conducted by Wang \textit{et al.} on AA$^{\prime}$ BL WSe${2}$. 
In the 1.0\%-OP case (Figs.~\ref{fig5}(d)), the material switches instead to a highly interesting state in which the minimum conduction band energies at K and $\Lambda$, as well as the minimum conduction band energies at K and $\Gamma$, align almost identically. Therefore, the KK, K$\Lambda$, $\Gamma$K , and $\Gamma\Lambda$ transitions become identical as 2.3 eV. 
Then, the ground-state exciton in the stacking-compressed case is also at $\mathbf{Q}=\boldsymbol{\Lambda}$, but this time it is of the type IDX$_{\textrm{HIL}}^{\Gamma\Lambda}$ at an energy of 1.536 eV. 
These results show the crucial influence of strain on the exciton energy spectrum of BL WSe$_2$.
The importance of strain is further compounded when switching on the external electric field, as significant differences in the field-dependent Stark shifts between the strained and unstrained BL AA$^\prime$ WSe$_2$ systems are apparent as evidenced in Fig.~\ref{fig6}, showing both distinct exciton energy alignments and nonlinear-to-linear Stark effect transitions (the DX$^{\mathrm{KK}}_\textrm{IL}$ shift always remaining fully linear, like in experiments, as the electron-hole layer separation of this $\mathbf{Q}=0$ exciton is not affected by the strain we applied, since it does not break any symmetries). 
With compressive strain, the $p$ and $t$ parameters do not change substantially (see third columng of Table~\ref{tab1}), the leading effect is the state-dependent rigid shift of the exciton energies. 
However, the $\mathbf{Q}=\mathbf{M}$ and  $\mathbf{Q}=\boldsymbol{\Lambda}$ involving transitions from the valence $d_{xy}+d_{x^2-y^2}$ orbitals at K/K$^{\prime}$ develop a plateau at low field, seemingly displaying more-than-quadratic dependence.
Instead, in the stacking-compressed case, the $t$-values reduce, indicating a tendency toward a more dipole-like behavior for the hybrid-IL excitons.

\section{Conclusions}
Our findings demonstrate how the linear and nonlinear Stark effect of bound electron-hole pairs in bilayer WSe$_2$ is chiefly determined by the distinct orbital and layer hybridization character of the underlying single-particle band structure.
We discuss the competition between dark, Stark-shifting states at different finite-momentum excitonic valleys together with the bright in-plane and interlayer states, resulting in internal field-dependent energy reordering within the bound exciton manifold, in agreement with existing experimental results.
We also discuss the sensitivity of the electronic excitations and the changes in the Stark effect due to mechanical strain on the samples.
The phenomenology we describe may be further experimentally probed via luminescence spectroscopy on carefully controlled samples. 
Our results may have broader applicability across other TMDs, although the relative energies of competing excitonic valleys may not be as closely matched as in the case of bilayer WSe$_2$. 
This fact, along with electric field and strain modulation, allows for the engineering of spatially separated excitons not only in terms of their energies but also in terms of the corresponding hybrid-interlayer character. 
In fact, the application of an out-of-plane electric field permits to distinguish exciton states originating from transitions between different regions of the reciprocal-space Brillouin zone (both direct and indirect processes). 
Moreover, the stacking and symmetry dependence of the field-mediated shift in exciton energies reveals a potential approach for identifying the stacking arrangement of TMD crystals.


\section{Acknowledgments}
This research was supported by Research Foundation - Flanders (FWO). The resources and services used in this work were provided by the VSC (Flemish Supercomputer Center), funded by the FWO and Flemish Government. 
This work was partially supported by ICSC - Centro Nazionale di Ricerca in High Performance Computing, Big Data and Quantum Computing -- funded by the European Union through the Italian Ministry of University and Research under PNRR M4C2I1.4 (Grant No. CN00000013).

\bibliography{achemso}

\end{document}